\documentclass[12pt]{article}
\usepackage{amssymb}
\usepackage{amsfonts}
\usepackage{graphicx}
\usepackage{psfrag}
\usepackage{epsfig}
\usepackage{latexsym}

\newcommand{\beq}{\begin{equation}}
\newcommand{\eeq}{\end{equation}}
\newcommand{\bea}{\begin{eqnarray}}
\newcommand{\eea}[1]{\label{#1}\end{eqnarray}}                                 \newcommand{\bdm}{\begin{displaymath}}
\newcommand{\edm}{\end{displaymath}}
\newcommand{\beqa}{\begin{eqnarray}}
\newcommand{\eeqa}{\end{eqnarray}}
\newcommand{\beqab}{\begin{eqnarray*}}
\newcommand{\eeqab}{\end{eqnarray*}}

 \def\@makefnmark{\hbox to 0pt{$^{\@thefnmark}$\hss}}  %ORIGINAL

\newcounter{saveeqn}%

\begin{document}

 %\hyphenation{}

% \begin{flushright}
% %NYU-TH/02-09-11\\
% hep-th/0304148
% \end{flushright}
 \vspace{0.5cm}
 \begin{center}
 {\Large \bf Regularization of Brane Induced Gravity}\\
 \vspace{0.7cm}
 %{\Large\bf Eventual Comment.}\\
 \vspace{0.5cm}
 {\large Marko Kolanovic \footnote{e-mail: mk679@nyu.edu}, 
  Massimo Porrati \footnote{e-mail:mp9@SCIRES.ACF.nyu.edu}, and\\
Jan-Willem Rombouts \footnote{e-mail: jwr218@nyu.edu}} \\
 \vspace{0.5cm}
 {\it Department of Physics, New York University,\\
 4 Washington Place, New York, NY 10003, USA.} \\
\date{\today}
\vspace{1.5cm}
 \end{center}

 \bigskip

% \begin{center}
% \bf Abstract 
\begin{abstract}
We study the regularization of theories of ``brane induced'' 
gravity in codimension $N>1$. 
The brane can be interpreted as a thin  dielectric with a 
large dielectric constant, embedded in a higher dimensional space.
The kinetic term for the 
higher dimensional graviton is enhanced over the brane.
A four  dimensional gravitation is found on the brane at 
distances smaller  than a critical distance $r<r_c$, 
and is due to the exchange of a massive resonant graviton. 
The crossover scale $r_c$  is determined by the mass of the resonance. 
The suppression of the couplings of light 
Kaluza-Klein modes to brane matter results in a 
higher dimensional force law at large distances. 
We show that the resulting theory is free of ghosts or tachyons.
\end{abstract}

% \end{center}

 \newpage

\section{Introduction}

Brane world  theories with large or infinite extra dimension provide 
insight into a number of problems of high energy physics, cosmology and 
possible relations between them~\cite{dgp,dgkn,SG,dvali,dgs}. 
Those models may ultimately emerge as a part of
a fundamental higher dimensional theory like string theory~\footnote{A first 
attempt towards realizing them in string theory was proposed in~\cite{amv}.}

The idea that our universe may have infinite extra dimensions was rejected 
for a long time as being incompatible with the observed 4D nature 
of gravity. Since gravity
probes all dimensions, the force between objects localized on a $3+1$ 
dimensional
hyper surface would be higher dimensional. Nevertheless, 
it was pointed out in~\cite{dgp}
that, in the presence of a brane in 5D space, the ordinary 5D 
action of gravity is modified by a 4D Einstein-Hilbert on the brane.
This term is compatible with all the symmetries of the theory and, therefore,
can be generated by quantum corrections. 
This model, also known as the  DGP (Dvali-Gabadadze-Porrati) model, 
is an example of a theory of ``brane induced gravity.'' There, gravity is
4D below a certain scale $r_c$, ranging from galactic to horizon size.  
 One major problem of this scenario, related to the existence of $1/r_c^2$ 
singularities in off-shell trilinear interactions of longitudinal 
gravitons~\cite{ddgv}, was pointed out in~\cite{a9} 
(see also~\cite{Rubakov}). There, it was shown that radiative corrections to 
DGP become uncontrollably large at a very low energy scale, 
$(L_{Pl} r_c^2)^{-1/3}$. This scale is essentially a cutoff beyond which the DGP
model needs a UV completion. A similar problem should arise also in codimension
greater than one, in analogy with the UV behavior of massive gravity~\cite{a0}.
We will not address this problem in this paper, but we will concentrate on 
another, simpler one. Namely, the proper definition and regularization 
of induced gravity in codimension $N>1$.

Setting aside the strong-coupling problem just mentioned here, 
brane induced gravity in five dimensions is compatible with
all present measurement and observations~\cite{dgkn} and allows the scale of
5D 
quantum gravity $M_*$ to be as low as $10^{-3}$eV~\cite{SG}. Choosing $r_c$ to
be the horizon radius, $M_*$ is instead $10^{8}$eV. In higher codimension, 
$M_*$ depends on the UV regularization scale, i.e. on the thickness of the 
brane. It is natural to choose this thickness to be $O(1/M_*)$; in this case 
$M_*$ becomes $O(10^{-3}\rm{eV})$. 

The investigation of brane induced gravity in a  space with more than one 
infinite extra dimension is interesting because  that framework may provide a 
solution to the cosmological constant problem \cite{dgs,w}.

Brane induced gravity in space with $N$ transverse 
dimensions is described by the action:

\beq\label{big}
S=M_{*}^{2+N}\int d^{4}xd^{N}Y\sqrt{G}{\cal R}_{4+N}+
M_{P}^2\int d^{4}xd^{N}Y\delta^{N}(Y)\sqrt{g}{\cal R}_{4},
\eeq
where $Y_{i},\quad i=1,..,N$ are coordinates of $N$ 
infinite extra dimensions, and the brane
is taken to be infinitesimally thin 
and located at the origin. ${\cal R}_{4+N}$ is the $(4+N)$D curvature,
$G_{MN}$ is the $(4+N)$D metric, ${\cal R}_4$ is the 4D curvature 
constructed from the induced metric
on the brane $g_{\mu\nu}\equiv G_{MN}(y=0)\delta^{M}_{\mu}\delta^{N}_{\nu}$. 
The philosophy behind  brane induced gravity is that there are two scales in 
the theory. A scale for bulk gravity $M_{*}$, 
and a much higher scale $M_{P}$, which characterizes the physics on the 
brane, i.e. the standard model  and possible extensions like SUSY and GUTs. 
In order to obtain the correct value of Newtons constant, $M_P$
is taken to be $\sim 10^{19}$GeV.

Calculating the Green function and KK spectrum of (\ref{big}) for the case 
$N=1$ is straightforward. The task is less clear for $N>1$ due to 
singularities that appear because of the zero thickness of the brane, and the
resulting singularities in the Green functions for $N>1$~\cite{sefu}. 

In this paper  we study a simple regularization of the action~Eq.(\ref{big}). 
We propose to consider an action for $(4+N)$-dimensional gravity, 
reminiscent of the action for a dielectric  in electrodynamics:
\beq
\label{sd}
 {S\over M_*^{2+N}}= 
\epsilon \int_{ \rho< \Delta } d^{4+N} x \sqrt{G_{4+N}}   
{\cal R}_{4+N}+ \int_{ \rho> \Delta}
 d^{4+N} x  \sqrt{G_{4+N}} {\cal R}_{4+N},
\eeq
with $ \epsilon$ a large dimensionless quantity (to be defined more precisely 
in the text), $\rho$ the radial coordinate in the $N$ transverse dimensions,
and $\Delta$ the width of the brane.

The brane in this theory is a thin ``gravitational'' dielectric.  
We will prove that, at distances smaller than a certain critical scale $r_c$ 
and for $N>1$, the 
gravitational interaction is mediated by a massive resonant graviton, whereas 
at larger distances one rediscovers $(4+N)$-dimensional gravitation.

We will also investigate smooth versions of this scenario. 
This is achieved by taking a varying ``dielectric constant.'' 
A scalar action of this kind was considered in~\cite{kol}, where it was 
pointed out that this kind of action can be realized physically in a 
dilaton gravity theory. 

We thus consider the action of \cite{kol} generalized to arbitrary codimensions:

\beq\label{act1}
{S\over M_{*}^{2+N}}=\int d^{4}xd^{N}Y\sqrt{G}\left[
1+\epsilon {\cal F}(\rho/\Delta)\right]{\cal R}_{4+N},
\eeq
where ${\cal F}(\rho / \Delta)$ is the shape of the brane, which we take to
depend only on the radial coordinate $\rho$ of the $N$ transverse dimensions.
Again, $\Delta$ is the width of the brane  and $\epsilon$ is a large, 
dimensionless parameter. 
The action Eq.~(\ref{sd}) is obtained from Eq.~(\ref{act1}) by
choosing for the profile the step function ${\cal F}=\theta(\rho-\Delta)$.

The action Eq.~({\ref{act1}) describes a smooth 3-brane in a 
$(4+N)$-dimensional space. 
We will show that for sufficiently large values of $\epsilon$, the effect of 
this brane ``dielectric'' constant is  to convert the higher dimensional laws 
of gravity to four-dimensional laws at short distances. 

On  physical grounds one does not expect that the introduction of a ``medium'' 
(i.e. a brane) into the higher dimensional vacuum will result in the 
appearance of ghosts
or tachyons and make the theory inconsistent. 
We will explicitly show this for the model (\ref{sd}). 
A brane with smoothly changing permeability, $\epsilon {\cal F}$, 
can be thought of as a succession of as a sequence of small regions of 
constant $\epsilon$, and hence the proof should also hold for smooth 
${\cal F}$. 

The action Eq.~(\ref{act1}) may  provide a physical realization of 
Eq.~(\ref{sd}). One can think of it as gravity coupled to a scalar that gets 
a non-constant vacuum expectation value, in the form of a topological defect 
(domain wall, string, monopole etc.).

This paper is organized as follows. 
In the first section we consider the scalar equivalent of action 
Eq.~(\ref{sd}). We analytically calculate the scalar field  resonances 
and show that they lead to the crossover distance between 4-dimensional and 
higher dimensional behavior for the force they mediate. 
We do that by studying the KK modes of the fields and prove that their 
coupling to the ``dielectric'' is determined by their wave function on the 
brane, for an appropriately chosen source. 
We will see that, generically, the coupling of heavy modes is strongly 
suppressed on the brane, and that the coupling of the modes is sharply peaked 
around a value, $m_g$, which is the mass of the effective 4-dimensional 
scalar ``graviton.'' 
 
Next, we take a specific choice of a smooth brane and see that even there
the scalar ``graviton'' is a massive resonance. Numerical results show that 
the behavior is the same as with the sharp-boundary dielectric. 
This shows that the regularization considered here is really shape 
independent. 

In Section 4, we investigate the tensor structure of our regularized theory. 
We show that in the model~(\ref{sd}) the 4 dimensional \textit{massive} 
graviton propagates with the correct degrees of freedom - i.e. the mass term 
generated in this theory has the Pauli-Fierz form. 
There are no ghost or tachyons propagating in our system, unlike in the
regularization proposed in~\cite{Dubovsky}. 

Section 5 summarizes the findings of this paper.

As we previously mentioned, its was pointed out in~\cite{a9} 
and~\cite{Rubakov} that the DGP theory suffers from strong-coupling problems, 
as is typical for theories with massive gravitons \cite{a0}. 
Our regularization, of course, does not cure that problem. 

%%%%%%%%%%%%%%%%%%%%%%%%%%%%%%%%%%%%%%%%%%%%%%%%%%%%%%%%%%%%%%%%%%%%%%%%%%%%%%%%%%%%%%%%%%%%%%%%%%%%%%%%

\section{A Dielectric Brane with a Sharp Boundary}
We start by looking at a model in which we imagine a thin 
``dielectric'' with a sharp boundary, and dielectric constant $\epsilon\gg 1$  
centered around the origin of the transverse $N$-dimensional space. 
 The results are
easily generalized to smooth branes (smooth variation of the ``dielectric'' 
constant) by taking that $\epsilon$ changes in small steps. 
 We are interested in whether this system exhibit a 4-dimensional 
behavior in a certain region of distance (or energy) scales. 

Let us analyze this system first for  the simplest case; a scalar field. 
Consider the following action ($M=0,...,3+N$):
\begin{equation}
{S\over M_*^{2+N}}= \epsilon \int_{ \rho < \Delta} d^{4+N} x 
(\partial_{M} \Phi)^2 + \int_{ \rho> \Delta} d^{4+N} x (\partial_{M} \Phi)^2,
\end{equation}
where we define the dimensionless quantity $ \epsilon $ as follows:
\begin{equation}\label{epsil}
\epsilon \Delta^N = M_P^2/M_*^{2+N},
\end{equation}
a definition guaranteeing that in the limit 
$ \Delta \rightarrow 0 $, $ \epsilon \Delta^N =\mbox{constant}$, 
we approach the delta function limit. $ \rho $ is the radial coordinate in the
extra N-dimensional space. 
To find a solution to this system, we solve the equation of motion in the
two different regions, and impose the matching conditions at the boundary of 
the sphere. One obtains:
\begin{eqnarray}
\Box \Phi^I  &=& 0, \quad \rho<\Delta, \\
\Box \Phi^o &=& 0,\quad \rho>\Delta,
\end{eqnarray}
and the boundary condition:
\bea
\Phi^I &=& \Phi^o \arrowvert_{ \rho=\Delta},\nonumber \\
 \epsilon \partial_\rho \Phi^I &=& \partial_\rho \Phi^o 
\arrowvert_{ \rho=\Delta}.
\eea{m1}
If we decompose the field in Kaluza-Klein modes, 
$ \Phi(x,Y)= \sum\sigma ^m(x) \phi_m(Y)$, 
$\Box_4 \sigma^m = m^2 \sigma^m$, the equations of motion and the boundary 
condition on the sphere reduce to equations for each one of the modes:
%\begin{eqnarray}
%\label{EOM}
%\Box_N \phi^o_m = -m^2 \phi^o_m, \\
%\Box_N \phi^I_m = -m^2 \phi^I_m. 
%\end{eqnarray}
\begin{equation}
\label{BC}
 \epsilon \partial_\rho \phi_m^I = 
\partial_\rho \phi_m^o \arrowvert_{ \rho=\Delta.}
\end{equation}
%The equations of motion become:
%\begin{equation}
%(\Box_4 \sigma^m) \phi_m   - \sigma^m (\Box_N  \phi_m) = 0.  
%\end{equation}
%Assuming $\Box_4 \sigma^m = -m^2 \sigma^m$ we get the equations of motions for the modes
% in each region to be of the Klein-Gordon form:
%\begin{eqnarray}
%\label{EOM}
%\Box_N \phi^o_m = -m^2 \phi^o_m, \\
%\Box_N \phi^I_m = -m^2 \phi^I_m. 
%\end{eqnarray}
We must make some important remarks at this point. 
We are interested in the coupling of the KK modes to the matter on the brane, 
which is given by the convolution of the wave function and the matter profile 
$\Psi(y)$.
\beq\label{convo1}
\tilde{\Phi}_{m}(0)=\int_{\rho<\Delta} d^{4+N}Y \Psi( Y)\Phi_{m}(Y).
\eeq
Now we can assume that  the profile of the matter fields is 
spherically symmetric in the extra dimensions, 
$\Psi(Y)=\Psi(\rho)$. We can take that the  standard model matter is given 
by zero modes of some fields with characteristic scale $M_{SM}\gg M_{*}$. 
Those zero modes have spherically symmetric profiles in the $N$ transverse
dimensions.
The excited states with nonzero angular momentum $l$ have masses which are 
naturally of order $M_{SM}$, so that they decouple from low energy physics. 
That is why we can assume that only KK modes with $l=0$  couple to the 
``spherically'' symmetric brane matter. When calculating the tree level
exchange, we are taking into account only a 
single tower of radial KK excitations.

Since we can assume that the matter profiles are sharply peaked around the 
origin, the coupling of the KK modes is approximately given by $\phi_m(0)$.

We look at the case of codimension 1 first. The DGP model is perfectly 
regular for $N=1$, 
so we expect to discover the same results as obtained in~\cite{dgp}.
%In this case, our results should be
%regularization independent for our construction to be compatible with DGP.
The general solution to the 5D equations of motion is 
(using plane wave normalization):
\begin{eqnarray}
\phi^I_m= A_{m}\cos(m \rho), \\
\phi^o_m=\cos(m \rho + \varphi). 
\end{eqnarray}
The odd modes do not couple significantly 
to matter located on the brane, as explained in the previous paragraph.
A straightforward calculation yields the following equation for $\varphi$:
\begin{equation}
 \epsilon \tan( m \Delta) = \tan( m \Delta + \varphi). 
\end{equation}
On the other hand $A_{m}$ is given by:
\begin{equation}
 A_m \cos( m \Delta) =\cos( m \Delta + \varphi).
\end{equation}
Since we are interested in four-dimensional distances larger than the width 
of the brane, modes
heavier than $1/\Delta$ decouple. 
Later we will take that $\Delta \sim 1/M_*$. Since $M_*$ is
the cutoff of the effective field theory for the bulk 
gravity, we take $ m \Delta \ll 1$. 
Then we can approximate $\tan( m \Delta)\approx m \Delta $,
so we find for $A_{m}$:
\begin{equation}
A_{m} \approx \frac{1}{\sqrt{1+(\epsilon m \Delta)^2}}.
\end{equation}
Using the definition of $ \epsilon $ (\ref{epsil}), we see that 
$\epsilon= \frac{r_c}{\Delta}$, where $r_c=\frac{M_P^2}{M_*^3}$. 
As explained above, the coupling of
the Kaluza Klein mode  of mass $m$ to the brane matter is given by:  
\begin{equation}
|\phi_{m}(\rho =0)|^2 \approx \frac{1}{1+(m r_c)^2}.
\end{equation} 
We see that we get a resonance, and that for
$m\gg r_c^{-1}$ the coupling is strongly suppressed. 
The result is independent of the
shape of the brane~\cite{kol} and it is exactly what is found in the 
DGP model~\cite{dgp}, as we anticipated. In ref~\cite{dgp} 
it was explained how this resonance leads to a $1/r$ potential for 
$r<r_c$ and to a $1/r^2$ potential for  $r>r_c$.

Now consider for instance the case of codimension $N=3$. 
The theory with a delta function type brane
is known to have singular behavior in this case.  Let us apply our 
regularization prescription to this case. 
We expect a regularization ($\Delta$)  dependent result in this case.
The solution in the two regions is:
\begin{eqnarray}
 \phi^I_m= A_{m} \frac{\sin(m \rho)}{\rho}, \nonumber\\
\phi^o_m=\frac{\sin(m \rho + \varphi)}{\rho}. 
\end{eqnarray}  
The $\phi^I_m$ solution is determined by asking for regularity at $\rho=0$.
We keep only the $l=0$ terms in the expansion in spherical harmonics as 
explained before.
Boundary conditions give the following equations:
\begin{eqnarray}
A_{m}\sin( m \Delta)=\sin(m \Delta + \varphi), \\
\epsilon [ m \Delta \cot(m \Delta) - 1] = m \Delta\cot( m \Delta + \varphi) -1.
\end{eqnarray}    
Expanding the cotangent, we get:
\begin{equation}
\cot( m \Delta + \varphi)\approx \frac{1}{m \Delta}-
\frac{\epsilon m \Delta}{3}.
\end{equation}  
We obtain for the amplitude:
\begin{equation}
 A_{m} \approx \frac{1}{ \sqrt{1- \frac{2 \epsilon (m \Delta)^2}{3}+ 
\frac{\epsilon^2}{9} (m \Delta)^4}}
\end{equation}
The amplitude squared is a sharply peaked function around its maximum, obtained
at $ \epsilon (m \Delta )^2 \approx 3 $. This gives, using our definition of
$\epsilon$, $m_g^2 \approx \frac{M_*^5 \Delta}{M_P^2}$.  
The coupling of the modes, given by the wavefunction squared, is now 
$\phi_m(0)^2 \approx m^2 A_m^2 $. We must emphasize that the fact that 
$A_m$ is singular at its maximum is due to our approximation. 
Indeed, the denominator receives corrections at $O[(m\Delta)^2]$, 
which gives $\phi_m(m_g)^2 \approx \frac{1}{\Delta^2}$: 
a finite and $\epsilon$-independent result.  For KK modes
with mass larger than $m_g$, the coupling to the brane is heavily suppressed. 
In this case we find that $m_g$  is $\Delta$ dependent, as 
we anticipated before. 

For $\Delta$, one might take the natural value $ \Delta=M_*^{-1}$, 
in which case the crossover distance and the graviton width become 
\beq\label{cros3}
r_{c}\equiv m_{g}^{-1}\sim \frac{M_P}{M_*^2},
\quad \Gamma_{m_g}\sim \frac{M_{*}^3}{M_{P}^2}.
\eeq
This is exactly what is found for the regularized delta-function type brane 
theory regularized
with higher derivative operators~\cite{sefu,Dubovsky}. 
In Fig.~1 we show the value of the wave function
$\phi_{m}(0)$ that is roughly the coupling of KK modes to the brane matter.

\vspace{5mm}
\centerline{\epsfig{file=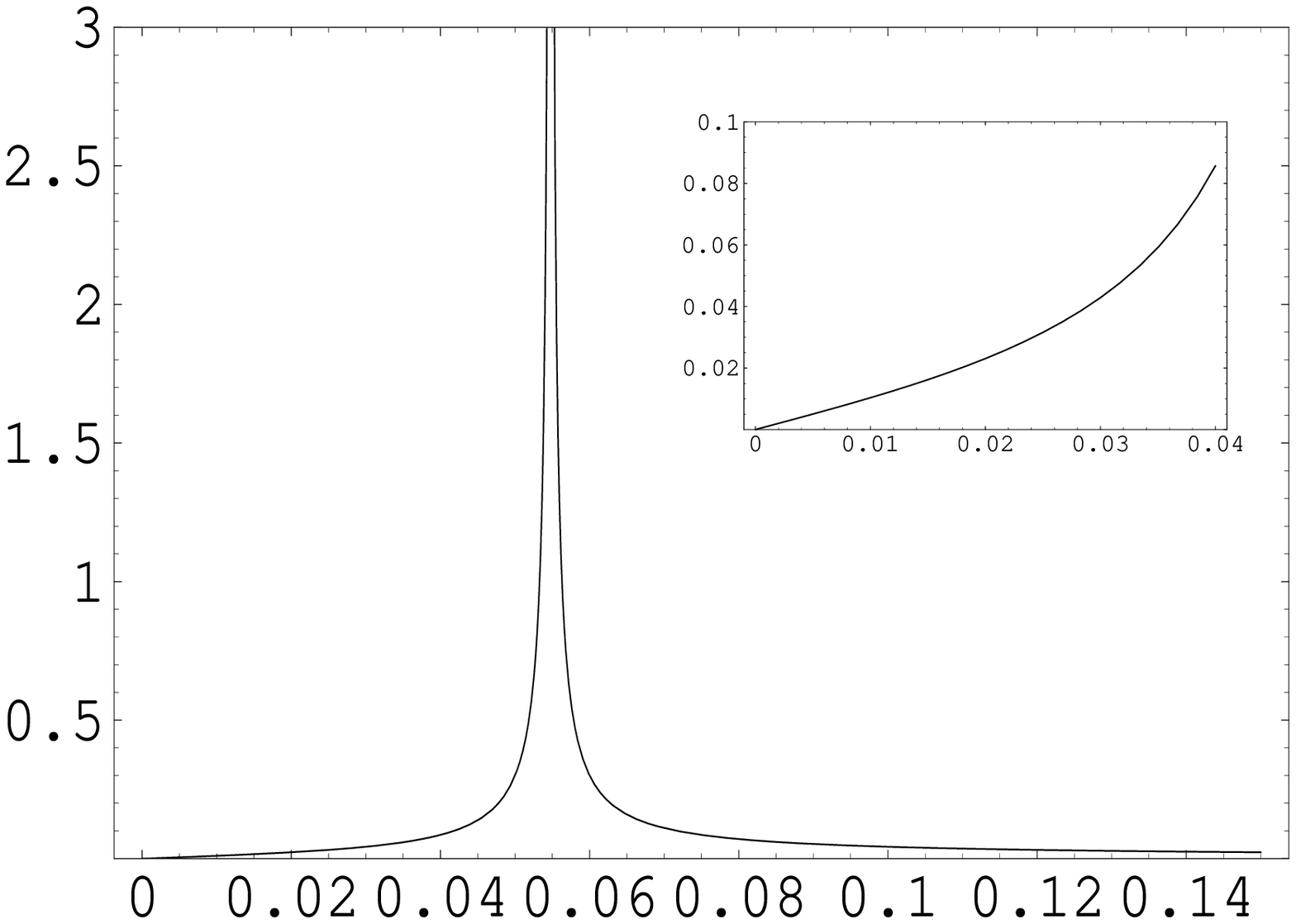,width=9cm}}
{\footnotesize\textbf{Figure 1:} Wave functions of KK modes on the brane 
for $\epsilon=1000$. 
The scalar ``graviton'' is a very sharp massive resonance. 
Inset: light modes are suppressed as  $\sim m$, giving rise
to a $1/r^4$ potential at large distances.}
\vspace{5mm}

%%%%%%%%%%%%%%%%%%%%%%%%%%%%%%%%

In codimension $N=3$, the  effective 4-D ``graviton'' 
is a very narrow resonance. The crossover from 4D to 7D behavior does not 
happen because of graviton ``decay.'' At large distances, the 
4D ``graviton'' simply becomes too heavy to contribute to the tree-level 
exchange. 

At large distances $r\gg r_{c}$ we evaluate the potential from the
coupling of light KK modes to brane matter. Their coupling is suppressed
linearly $\phi_{m}(0)\sim m$ and the potential is:

\beq\label{7dpo}
V_{New}(r)\approx \int_{0}^{\infty}\frac{m^2}{M_{*}^{5}}
\frac{\exp(-mr)}{r}dm\sim\frac{1}{M_{*}^5}\frac{1}{r^4}
,\quad\quad r\gg r_{c}.
\eeq
This is just the 7D potential that appear at distances larger than 
the crossover distance.

At short distances, the Newtonian potential  between two masses,  
due to exchange of KK modes, $\phi_m$, is dominated by the 
exchange of the lowest-mass resonance. 
So we can approximate our integral by the area of the resonance peak.
This gives:
\beq\label{7dpoa}
V_{New}(r)\approx \int_{m_g-\Gamma_{m_g}}^{m_g+\Gamma_{m_g}}
\frac{|\phi_{m}(0)|^{2}}{M_{*}^{5}}\frac{\exp(-mr)}{r}dm \sim  \frac{1}{M_P^2} \frac{1}{r}
\quad r\ll r_{c}.
\eeq
We see that the weak four dimensional gravity is due to the exchange of a 
massive resonant ``graviton.''

%%%%%%%%%%%%%%%%%%%%%%%%%%%%%%%%

\section{A Smooth  Dielectric Brane}

In this section we study the 4D effective field theory of a scalar system in 
which the kinetic term of a bulk field is not homogeneous through extra 
dimensions. We shall obtain results qualitatively similar to the ones obtained
in the previous Section. As we remarked earlier, a smooth ``dielectric'' 
constant may be achieved in dilaton  gravity theories. The reader not
interested in the details of this regularization can skip this section and 
proceed to Section 4, where the sharp-boundary approximation is applied to
true spin-2 gravity.

The action that describes a scalar in a smoothly varying ``dielectric'' medium is

\beq\label{firstl}
{\cal L}=M_*^{2+N} f(Y_{i})\partial_{A}\Phi\partial^{A}\Phi.
\eeq
We take  the localizing profile $f(Y_{i})$ to be a spherically
symmetric function of radial variable in $N$ extra dimensions. 
For the system~Eq.(\ref{act1})
that we are studying the localizing profile $f$ is

\beq\label{f}
f(\rho/\Delta)=1+\epsilon {\cal F}(\rho/\Delta).
\eeq
As before, we are interested in the four dimensional spectrum of 
Eq.~(\ref{firstl}), and we decompose the field $\Phi$
into its Kaluza Klein modes
\beq\label{m2} 
\Phi(x_{\mu},Y_{i})=\sum_{m}\Phi_{m}(Y_{i})\sigma_{m}(x_{\mu}), 
\qquad \Box_{4}\sigma_{m}=m^2\sigma_{m}.
\eeq
The differential equation for wave functions $\Phi_{m}(Y_{i})$ is

\beq\label{ekv}
f(\rho)\nabla_{N}^{2}\Phi_{m}+\nabla_{N} f(\rho)\nabla_{N} 
\Phi_{m}+f(\rho)m^2\Phi_{m}=0.
\eeq
Let us look at the Schr\"odinger equation for wave functions $\Phi_{m}$ 
for a general profile $f(\rho)$ in the case $N=2$. 
The Laplacian and gradient in $N=2$ have both radial and angular
parts, so the solutions can be separated as 
$\Phi_{m}(\rho,\theta)=\sum\Phi_{n}(\rho)\Theta_{m}(\theta)$, 
where $\theta$ is the polar angle in the transverse plane.
The scalar ``graviton'' couples to matter located on the string-like brane 
$\epsilon{\cal F}(\rho)$.  We assume, as explained in the previous section, 
that only the radial KK modes
(no angular dependence) of our scalar field 
couple to matter density localized on the soliton $\Psi(\rho)T(x)$.
If we substitute

\beq\label{subs}
\Phi_{m}(\rho)\equiv\frac{1}{\sqrt{\rho f(\rho)}}\phi_{m}(\rho),
\eeq
we obtain a Schrodinger equation for modes $\phi_{m}$

\beq\label{s2}
-\phi^{''}_{m}+\left(\frac{1}{2}\frac{f''}{f}+
\frac{1}{2}\frac{f'}{f}\left(\frac{1}{\rho}-\frac{f'}{2f}\right)^2-
\frac{1}{4\rho^2}\right)\phi_{m}=m^2\phi_{m}.
\eeq
The modes are orthogonal, and the normalization integrals are:

\beq\label{orth}
\int f(\rho)\Phi_{n}\Phi_{m}\rho d\rho=\alpha_{m}\delta_{m,n},\quad
\frac{\int f(\rho)\partial_{\rho}\Phi_{n}\partial_{\rho}\Phi_{m}\rho d\rho}
{\int f(\rho)\Phi_{n}\Phi_{m}\rho d\rho}=m^2.
\eeq
The effective 4D action is:

\beq\label{efac}
{\cal L}=\frac{1}{2}M_*^{2+N}
\sum_{m}\left( \partial_{\mu}\sigma_{m}\partial^{\mu}
\sigma_{m}-m^2\sigma_{m}^{2}\right)+\sum_{m}\frac{\tilde{\Phi}_{m}(0)}
{M_{*}^2\sqrt{\alpha_{m}}}\sigma_{m}T(x).
\eeq
Here $\tilde{\Phi}_{m}(0)$ is the convolution of the wave function $\Phi_{m}(\rho)$ and the
matter profile $\Psi(\rho)$

\beq\label{convo2}
\tilde{\Phi}_{m}(0)=\int \rho d\rho\Psi(\rho)\Phi_{m}(\rho).
\eeq

As in the previous section, this represents the coupling of KK graviton modes 
of mass $m$ to brane-localized matter, and we can approximate their couplings 
to  be $\tilde{\Phi}_{m}(0)\approx \Phi_{m}(0)$.

Let us illustrate the main properties of the smooth brane  model in $N=2$ 
transverse dimensions. Consider following profile:

\beq\label{delta}
{\cal F}=\frac{\exp{(-\rho^2/\Delta^2})}{\rho/\Delta}.
\eeq
This profile, multiplied by $1/\Delta^2$ is a regularization of a
delta function in $N=2$ dimensions. 
It is instructive to see what the spectrum is when one neglects the constant 
1 in Eq.~(\ref{f}), and then treats the problem in some approximation. 
This is equivalent to taking 
the limit $\epsilon \longrightarrow \infty$. Taking $f=1+\epsilon {\cal F}$ one finds in Eq.~(\ref{s2}) the
Schr\"odinger potential

\beqa\label{sho}
V(\rho)=(1/\Delta)^2[(\rho/\Delta)^2-1].
\eeqa
This is the potential of a simple harmonic oscillator with the spectrum 

\beq\label{spho}
m_{n}=\sqrt{2n}/\Delta,\quad n=0,2,4,...
\eeq
Here $n$ is an even integer, 
since the wave function and derivative at origin must
be continuous, and we restricted ourselves to modes with no angular dependence.
As a first approximation, we can say that the spectrum of the model in 
Eq.~(\ref{act1})
consists of a tower of metastable modes with masses Eq.~(\ref{spho}). 
A metastable zero mode (with mass exactly zero in the 
$\epsilon\longrightarrow \infty$ limit) will be responsible for the $1/r$
potential\footnote{In fact, in the case $N=1$ \cite{kol} this 
is an exact picture. The metastable mode of mass zero has a width 
$\sim 1/\epsilon$, that results in the crossover distance 
$r_c=M_{P}^2/M_{*}^3$ between four and five-dimensional gravity.}.
The Schrodinger potential resulting from the profile (\ref{f}) reads

\beqa\label{sh23}
V(\rho)=(1/\Delta)^2\frac{\exp(2\rho^2)-4\epsilon[2\rho^3\exp(\rho^2)+\epsilon
(-1+\rho^2)]}{4[\epsilon+\rho\exp(\rho^2)]^2}.
\eeqa
where $\rho$ is given in units of the brane 
thickness $\Delta$. This potential is shown in Fig.~2. 
The tunneling rate 
represents the decay width of the resonance and can be calculated in the 
WKB approximation

\beq\label{decw}
\Gamma_{0}\sim \frac{1}{\Delta}\exp\left( -2\int_{y_1}^{y_2}\sqrt{V}dy\right),
\eeq
here $y_1,y_2$ are the classical turning points. The width of the 
potential well is $\sim \Delta$. 
For large values of $\epsilon$, we are well within the limit of validity of
the WKB approximation, namely $V'/(2V)^{3/2}\ll 1$.  
The integral in Eq.~(\ref{decw}) can be evaluated numerically. 
The dependence of the integral on $\epsilon$ is (shown in a log-log plot in 
the insets in Fig.~2) $\Gamma^{N=2}_{0}\sim 1/\Delta\epsilon^{0.931}$.

\vspace{5mm}
\centerline{\epsfig{file=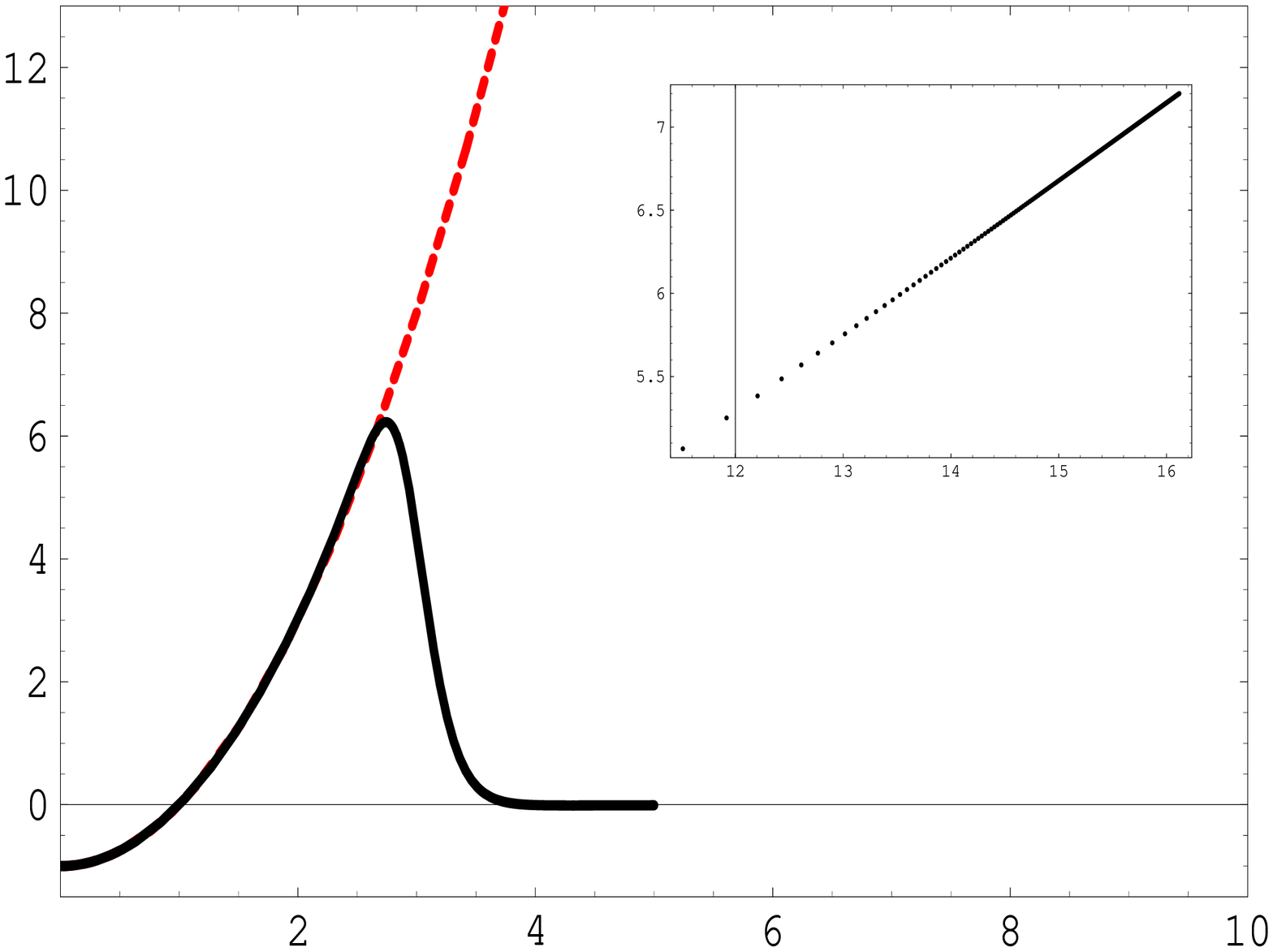,width=7cm}}
{\footnotesize\textbf{Figure 2:} 
Potential Eq.~(\ref{sh23}) $N=2$ (solid line) together with potential
Eq.~(\ref{sho}) (dashed line). $\rho$ is in units of $\Delta$, the potentials 
are in units of $1/\Delta^{2}$,
while $\epsilon=10000$. The inset shows the log-log plot of  
integral in Eq.~(\ref{decw}) vs. $\epsilon$.}

Here, the difference from the expected result $\Gamma\sim 1/\epsilon$
is due to the WKB approximation. Assuming that the width of zero mode is 
$\sim 1/\epsilon$, and by
knowing that the zero mode is responsible for the 4D potential on the brane,
we can calculate the value of the ``Newton'' constant $G$. 
The square of KK graviton coupling to matter (\ref{efac})
multiplied by the width of the zero mode resonance gives
the cumulative effect of tree level exchange of continuum KK modes in the 
resonance.
The effective 4D Newton constant is thus 
$G_{N=2}\sim\frac{1}{M_{P}^2}$ for two choices of parameters: 1)
$\Delta\sim 1/M_*$ and $\epsilon \sim (M_P/M_*)^2$; 2) 
$\Delta\sim 1/M_P$, and $\epsilon \sim (M_P/M_*)^4$.

Four dimensional interactions between masses on the brane are mediated by the 
exchange of the
zero mode resonance, of width $\sim 1/\Delta\epsilon$.
The WKB approximation gives the value of Newton's constant, 
however it misses most of the
features of brane induced gravity in $N=2$ transverse dimensions. 
In particular, it doesn't predict the transition between the four-dimensional 
and  the higher-dimensional regime and it does not give
the distance at which this transition happens (crossover distance). 
In the rest of this section, we will study the spectrum and the couplings
of KK modes in brane induced gravity. 
We will see that the graviton resonance has a finite mass, and that its width 
is not $\sim 1/\epsilon$. However, the correct value of the Newton constant
is obtained by exchange of a massive graviton resonance.

The Schroedinger equation can be solved numerically, and its solution can be 
used to find the convolution of the KK wave functions with the wave function
of  localized matter. We will investigate the couplings of different
modes in order to determine how the transition between four dimensional gravity
and higher dimensional gravity occurs and what is the crossover distance. 

The suppression of the KK graviton couplings to brane matter is shown in 
Fig.~3. We notice the peak positioned
close to zero mass that is responsible for 4D gravity, as well as higher 
resonant modes at position of harmonic oscillator levels, 
with masses $m=2/\Delta,\sqrt{8}/\Delta, etc.$. Since we are interested
in the large distance behavior, we show a 
magnified picture of a graviton resonance in log-log scale. 
The behavior is as follows: for the mode of mass zero, the coupling is zero. 
Then, the coupling rises as
$\sim \sqrt{m}$, with the peak positioned at 
$m_{g}\sim 1/\Delta\sqrt{\epsilon}$. To the right of the peak, the 
coupling dies off as $\sim 1/m$.

\vspace{5mm}
\centerline{\epsfig{file=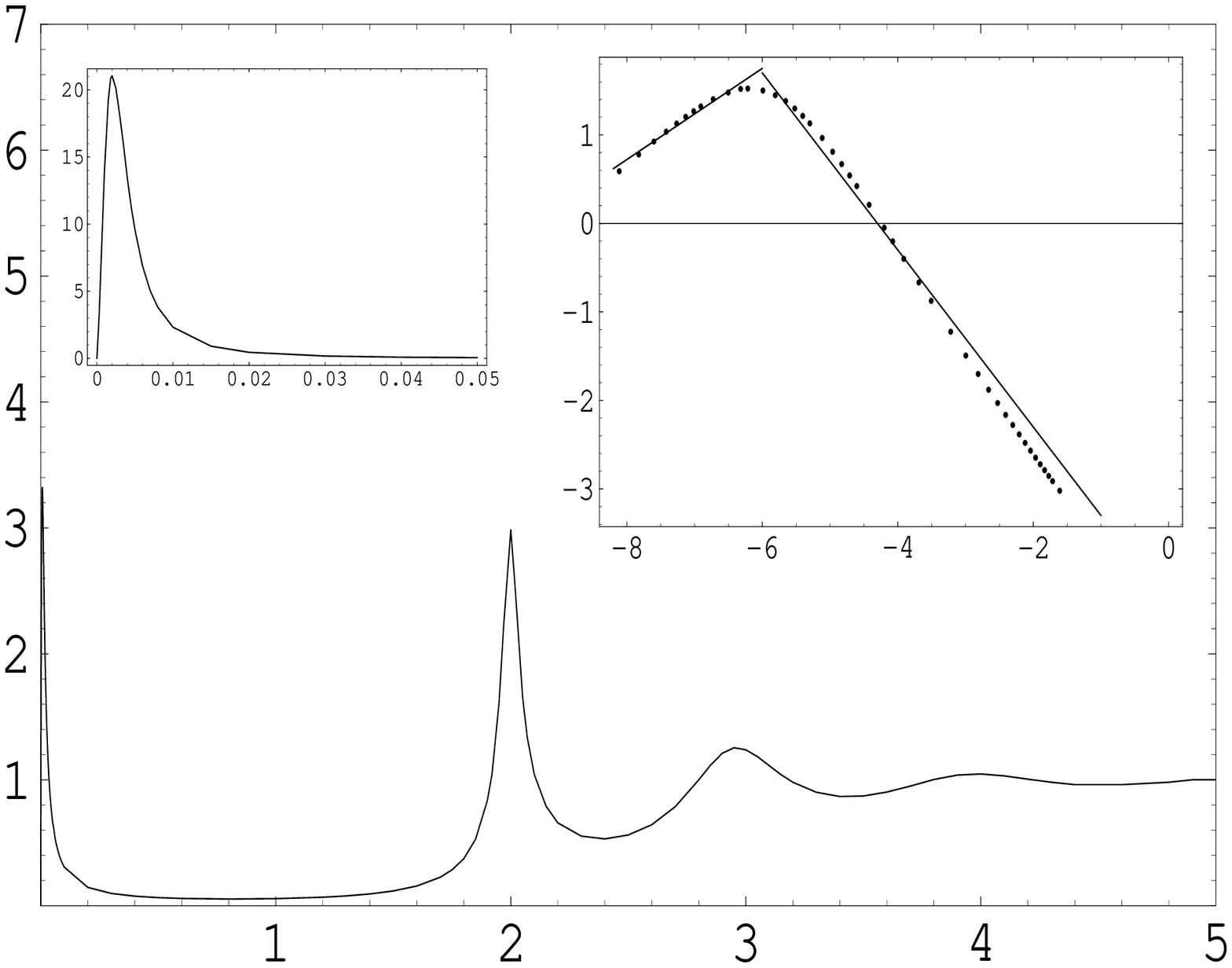,width=9cm}}
{\footnotesize\textbf{Figure 3:} Dependence of the coupling of a 
KK mode on its mass for $\epsilon=5000$. 
Peaks are at the location of resonant states of a harmonic oscillator with
$m=0,2,\sqrt{8}...$.
In the inset, a peak is shown at $m_{g}\sim 1/\Delta\sqrt{\epsilon}$, 
for $\epsilon=50000$ both in normal (left) and log-log (right) scale. 
The ``graviton'' resonance is asymmetrical. To the left of the peak,  
it grows as $\sim \sqrt{m}$, giving rise
to a 6D potential at large distances, while to the right, it decays as 
$\sim 1/m$. The solid lines have coefficients 0.5 and -1, respectively.}
\vspace{5mm}

We see that the graviton is a massive asymmetric resonance. 
The mass and the width of the resonance are

\beq\label{masg}
m_{g}\sim \frac{1}{\Delta\sqrt{\epsilon}},\quad \Gamma_{m_g}\sim\frac{1}{\Delta\sqrt{\epsilon}}.
\eeq
Starting from zero mass, the resonance grows as 

\beq\label{gro}
\phi_{m}(0)\sim \sqrt{\epsilon m}.
\eeq

Now we can choose two possible sets of values for 
$\epsilon$ and $\Delta$ in order to
reproduce the correct 4D Newton constant $\Delta\sim 1/M_*, \epsilon 
\sim (M_P/M_*)^2$; or 
$\Delta\sim 1/M_P, \epsilon \sim (M_P/M_*)^4$. Both choices give,
for the crossover distance (inverse graviton mass) and the resonance width,

\beq\label{cros2}
r_{c}^{N=2}\equiv m_{g}^{-1}\sim \frac{M_P}{M_{*}^2},
\quad \Gamma_{m_g}\sim\frac{M_{*}^2}{M_{P}}.
\eeq
In the case of $N=2$, the graviton mass is roughly the same as the resonance  
width.
On the brane, at distances $r\gg r_c$, we can evaluate the ``Newtonian'' potential
$V_{New}$ from the coupling  given in Eq.~(\ref{efac}), and the shape of the peak's
left ridge Eq.~(\ref{gro}). The exchange of modes with mass $m>1/r$
contributes weakly, so one can integrate from zero to infinity. 
The coupling of graviton KK modes
is proportional to $\sim\sqrt{m}$, so the potential is given by:

\beq\label{6dpo}
V(r)\approx \int_{0}^{\infty}\frac{m}{M_{*}^{4}}
\frac{\exp(-mr)}{r}dm=\frac{1}{M_{*}^4}\frac{1}{r^3}
,\quad\quad r\gg M_{P}/M_{*}^2.
\eeq
This is just the 6D Newton's potential that appear at distances larger than 
the crossover distance.
At short distances, the exchange of a single massive-graviton resonance 
gives rise to the 4D Newton's potential: 

\beq\label{6dpoa}
V(r)\approx \frac{1}{M_{P}^2}\frac{1}{r},\quad
\quad r\ll M_{P}/M_{*}^2.
\eeq
Here we approximated the area under the square of the curve determining the
coupling with twice the area on the left side of the peak.

%%%%%%%%%%%%%%%%%%%%%%%%%%%%%%%%%%%%%%%%%%%%%%%%%%%%%%%%%%%%%%%%%%%%%%%%%%%%%%%%%%%%%%%%%%%%%%%%%%%%%%%%

\section{The Spin 2 Case}
After having studied in details the scalar theory, we can proceed 
with a true gravitational theory. 
Although the scalar gravity example captures the main features of 
brane-induced gravity,
it is crucial to show that our regularization gives a consistent (4D) 
spin-2 theory.
In particular, we have to show that no tachyons
or ghosts propagate in our framework. 
In the effective 4D theory, the higher dimensional graviton is 
represented by  four dimensional spin two states and a set of scalars. 
We should check that the masses given to those fields are not
negative, and that the (massive) graviton propagates with the right number 
of degrees of freedom, which is assured if the mass term is of the 
Pauli-Fierz form. 

The action to be considered is:

\begin{equation}
{S\over M_*^{2+N}}= \epsilon \int_{ \rho< \Delta } d^{4+N} x 
\sqrt{G_{4+N}}   {\cal R}_{4+N}+ \int_{ \rho> \Delta}
 d^{4+N} x  \sqrt{G_{4+N}} {\cal R}_{4+N}.
\end{equation}

We can do an expansion in KK-modes for the gravitational field, 
analogous to the scalar case:

\begin{equation}
 G_{MN}=\sum_{m} G^m_{MN}(x) \phi_m(Y). 
\end{equation}
The boundary conditions now are the well-known 
Israel matching conditions~\cite{Israel:rt}, which give an equation for
the extrinsic curvature of the sphere which divides the 
$4+N$-dimensional space into two parts:

\begin{equation}
 \epsilon K_{MN}^I (x,Y)= K_{MN}^o (x,Y)\vert_{\rho=\Delta}. 
\end{equation}
These equations become particularly simple when we use 
Gaussian normal coordinates, since then
the extrinsic curvature is given by 
$K_{MN}=\frac{1}{2} \frac{ \partial G_{MN}}{\partial \rho}$.
One sees that the boundary conditions for the KK modes are completely 
analogous to the scalar case
equations treated in the previous section. This reduces the spin 2 case
essentially to the spin 0 case. 
In the following  equations we split the $4+N$-dimensional indices $M,N, ...$ into
4D indices: $\mu=0,...,3$, and transverse indices: 
$i=4,...,3+N$. The $\gamma_{ab}$ are the fluctuations
of the metric around the flat background and 
$\gamma=\gamma_4+\gamma_N$ is the trace of the fluctuation. The Einstein 
tensor is denoted by ${\cal G}_{MN}$.

\bea
{\cal G}_{i \mu}&=&\frac{1}{2} \partial^\lambda \partial_i 
\gamma_{\mu \lambda}+\frac{1}{2} \partial^j  \partial_i \gamma_{\mu j}+
\frac{1}{2} \partial^\lambda \partial_{\mu}\gamma_{i \lambda}+\frac{1}{2} 
\partial^j \partial_{\mu} \gamma_{i j} \nonumber\\
&& -\frac{1}{2}\Box_4 \gamma_{i \mu} 
-\frac{1}{2}\Box_N  \gamma_{i \mu}-
\frac{1}{2} \partial_{\mu} \partial_i \gamma,
\eea{m3}

\bea
{\cal G}_{\mu \nu} &=& 
\partial^\lambda \partial_{( \mu} \gamma_{ \nu) \lambda}-\frac{1}{2}\Box_4 
\gamma_{ \mu \nu}-\frac{1}{2} \partial_{\mu} \partial_{\nu}
( \gamma_4+\gamma_N)+\partial^i\partial_{( \mu} \gamma_{ \nu) i}-
\frac{1}{2}\Box_N \gamma_{ \mu \nu} \nonumber\\ &&
-\frac{1}{2}\eta_{\mu \nu}\lbrack \partial^\lambda \partial^\rho 
\gamma_{\lambda \rho}-(\Box_4+\Box_N)\gamma_4 \rbrack \nonumber \\ &&
-\frac{1}{2}\eta_{\mu \nu}\lbrack \partial^i \partial^j \gamma_{i j}+
2 \partial^i \partial^{\mu} \gamma_{i \mu}-(\Box_4+\Box_N)\gamma_N \rbrack.
\eea{m4}
We perform now a shift on $\gamma_{\mu \nu}$:

\beq{\label{mm}}
 \gamma_{\mu \nu} \rightarrow \gamma_{\mu \nu}+ \alpha \eta_{\mu \nu} 
\gamma_N. 
\eeq
After some algebra we get:

\bea
{\cal G}_{\mu \nu}&=& \partial^\lambda \partial_{( \mu} \gamma_{ \nu) \lambda}
-\frac{1}{2}\Box_4
 \gamma_{ \mu \nu}-\frac{1}{2} \partial_{\mu} \partial_{\nu} \gamma_4+
\partial^i\partial_{( \mu} \gamma_{ \nu) i}-\frac{1}{2}\Box_N 
\gamma_{ \mu \nu} \nonumber\\
&& -\frac{1}{2}\eta_{\mu \nu}\lbrack \partial^\lambda \partial^\rho 
\gamma_{\lambda \rho}-(\Box_4+\Box_N)\gamma_4 \rbrack \nonumber \\ &&
-\frac{1}{2}\eta_{\mu \nu}\lbrack \partial^i \partial^j \gamma_{i j}
-(\Box_4+\Box_N)\gamma_N \rbrack +\frac{1}{2} \alpha \eta_{\mu \nu} \Box_N \gamma_N.
\eea{m5}
Now we make the gauge choice:

\begin{equation}
 \partial^i \gamma_{ \nu i}=0, \qquad
\partial^i \partial^j \gamma_{i j} -\alpha \Box_N \gamma_N=0.
\end{equation}
For $\alpha=-\frac{1}{2}$ we get the following equation of motion for 
the four dimensional part of the metric:
\begin{equation}
\label{G^4}
{\cal G}^4_{\mu \nu}-\frac{1}{2} \Box_N \gamma_{\mu \nu}+\frac{1}{2} 
\eta_{\mu \nu} \Box_N \gamma_4 =T_{\mu \nu}.
\end{equation}
We assume that the matter distribution is confined to the brane, so only 
the 4D part of the energy momentum tensor is nonzero. 
If we contract this equation with $\partial^{\mu}$, we get, by conservation of
 the matter energy-momentum tensor, and thanks to the Bianchi identities:
\begin{equation}
 \Box_N( \partial^\mu \gamma_{ \mu \nu}-\partial_{\nu} \gamma_4)=0.
\end{equation}
This equation gives us a constraint on $\gamma_{\mu \nu}$. Indeed, 
remembering that 
$\Box_N$ applied to the fields gives their (nonzero) masses, we can conclude:
\begin{equation}
 \partial^\mu \gamma_{ \mu \nu}-\partial_{\nu} \gamma_4=0.
\end{equation}
On the other hand, if we take the trace of Eq.~(\ref{G^4}), we get 
\begin{equation}
\label{T}
 \frac{3}{2} \Box_N \gamma_4=T.
\end{equation}
This means that $\gamma_4$ is determined ``algebraically'' by T; 
$\gamma_4=\frac{2}{3 \Box_N} T$. By algebraically, we mean that the equation 
for $T$ is local in four dimensions.
Next, after a short calculation, 
the equations of motion for $\gamma_{\mu i}$ become: 
\begin{equation}
  \frac{1}{2} \partial^\lambda F_{\lambda \mu}^i+\frac{1}{2} 
\Box_N \gamma_{\mu i}=0,
\end{equation}
where $F^i_{\mu \nu}$ is the field strength of the 4-dimensional vectors 
$\gamma_{\mu i}$.
This equation describes $N$ massive vector fields in four dimensions. 

We are left with the equations of motion of the $N^2$ scalars $\gamma_{ij}$. 
One obtains:
\begin{equation}
 \Box_4 \gamma_{ij}+\Box_N \gamma_{ij}+\partial_i \partial_j \gamma+ \eta_{i j}
(\frac{1}{2} \Box_4 \gamma_N+\frac{1}{2} \Box_N \gamma_N-\Box_N \gamma_4)=0.
\end{equation}
Observe that the divergence of ${\cal G}_{ij}$ is identically zero by 
virtue of the gauge choice $\partial^j \gamma_{ij}+\frac{1}{2} \partial_i 
\gamma_N=0$\footnote{As it should be, using global energy-momentum 
conservation and remembering that the energy momentum tensor has only 4D 
components.}. 

Using Eq.~(\ref{T})  we finally obtain:
\begin{equation}\label{Gij}
 \Box_4(\gamma_{ij}+\frac{1}{2} \eta_{ij} \gamma_N)+
\Box_N(\gamma_{ij}+\frac{1}{2}\eta_{ij} \gamma_N)=(\eta_{ij} 
\Box_N - \partial_i \partial_j)\frac{2}{3 \Box_N} T.
\end{equation}

Looking at equations (\ref{G^4}) and  (\ref{Gij}), we see that all the mass 
terms have the correct form, with $m^2=-\Box_N$. In particular, the spin-2
mass terms have the Pauli-Fierz form. This is of course what we expect
in a generic Kaluza-Klein reduction~\cite{duff}. We can thus 
be sure that in our regularization, no ghost propagates. 

\section{Conclusions}

In this paper we have proposed a well-defined regularization scheme for 
``brane-induced'' gravity in codimension $N>1$. At its simplest, the scheme
replaces an infinitely thin brane with a ``dielectric'' sphere of radius 
$\Delta$ in the transverse space. This regularization allows for a simple
analytic computation of the spectrum of massive resonances, and it accounts 
simply for the tensor structure of gravity. Unlike other regularizations
already existing in the literature, ours introduces neither ghosts nor 
tachyons. 
In this respect we must notice an important point. 
Our definition of the 4D graviton involves a shift [see Eq.~(\ref{mm})], which
is the linearized version of a conformal rescaling. By doing this shift, we
couple the (scalar) internal components of the metric fluctuation to the trace
of the matter stress-energy tensor, even though it only has 4D components.
In other words, extra scalar, dilaton-like degrees of freedom couple to matter.
This is not a disaster {\em per se}, because even in their absence, linearized
massive gravity propagates an extra spin-zero degree of freedom~\cite{vdvz}.
Their presence only worsens the vDVZ discontinuity. 
In the regularization of ref.~\cite{Dubovsky}, instead, when the regularization
parameter is removed, one recovers {\em massless} 4D gravity. This is achieved
at the price of introducing light ghosts into the system.

We notice also that in codimension $N=1$, no extra scalar propagate, as shown 
by Eq.~(\ref{Gij}).

Finally, our ``sharp'' regularization can be easily extended to cover the case
of vector fields in $4+N$ dimensions. The smooth version of our regularization
may also be realized in a physical setting by coupling $4+N$ dimensional gravity to
a scalar field which admits a soliton solution (kink etc.)
\subsection*{Acknowledgments}
M.K. would like to thank G. Dvali for very useful communications.
J.-W.R. would like to thank F. Nitti for useful discussions. 
M.P. is supported in part by NSF grant PHY-0070787.

\end{document}